\documentclass[]{JHEP3} % 10pt is ignored!
\usepackage{epsfig,multicol,bbm}

\title{A Trouble with Ho\v{r}ava-Lifshitz Gravity}
\author{Miao Li, Yi Pang\\Key Laboratory
of Frontiers in Theoretical Physics,\\Institute of Theoretical
Physics, Chinese Academy of Sciences, Beijing 100190, P.R.China
    \\
    E-mail: \email{mli@itp.ac.cn}\\E-mail: \email{yipang@itp.ac.cn}}

\preprint{CAS-KITPC/ITP-112}

\abstract{We study the structure of the phase space in
Ho\v{r}ava-Lifshitz theory. With the constraints derived from the
action, the phase space is described by five fields, thus there is a
lack of canonical structure. The Poisson brackets of the Hamiltonian
density do not form a closed structure, resulting in many new
constraints. Taking these new constraints into account, it appears
that there is no degree of freedom left, or the phase space is
reduced to one with an odd number of fields.}

\begin{document}

\section{Introduction}

Ho\v{r}ava recently proposed a gravity theory with asymmetry between time and
space \cite{Horava1,Horava2}. This theory is non-relativistic in the UV limit,
thus it is hoped that it is UV finite. It is similar to a scalar field theory
of Lifshitz \cite{Lifshitz} in which the time dimension has weight 3 if a space dimension has weight 1, thus this
theory is called Ho\v{r}ava-Lifshitz gravity. Ho\v{r}ava argued that it is superficially
renormalizable based on power counting and may flow to Einstein's general relativity in the IR region.
This work has stirred up a surge of research on possible applications of the theory
to cosmology and black hole physics. We have no intention to be complete in offering the literature,
for those interested in a list of these papers, we refer to a most recent paper \cite{Chen}.

When any theory claims to be a renormalizable field theory of
gravity, one must exercise great care, especially with a theory
without general covariance to begin with, since we know that general
covariance is hard to avoid in a theory with a massless spin 2
particle. One may already poses questions at face value. If a theory
can flow to Einstein's theory only when a cosmological constant is
introduced, how can one avoid this cosmological
constant?\footnote{Some papers also pointed out the problem related
to cosmological constant \cite{Nastase, Visser}.} If one fine-tunes
parameter $\lambda$ in the kinetic term in the Ho\v{r}ava action to
1/3 thus makes the cosmological constant vanish, then how $\lambda$
can flow to 1 in Einstein theory? And, how a field theory of gravity
can explain the fact that the maximal entropy of a region is
proportional to the area of the surface surrounding it?

With the above questions in mind, we begin a study on constraints in
this theory. One easily comes to doubt that whether the system of
constraints of the theory makes sense, since not all constraints
correspond to local symmetries. Ho\v{r}ava retains all
diffeomorphism symmetries in space, but gives up on time local
symmetry. In a constrained system, constraints are normally
generators of symmetries, thus they are guaranteed to form a closed
system under the Poisson bracket. Now, the constraints corresponding
to the lapse function have no corresponding symmetries, it is
natural that they will generate new constraints under the Poisson
bracket. Indeed they do, as we will show shortly.

One may simply choose not to impose these new constraints. If so, then there are
four families of constraints: one corresponding to the lapse function $N$, three corresponding
to shift functions $N^i$. In the ADM canonical formalism (especially suitable for the
non-relativistic theory of Ho\v{r}ava), there are 12 fields on the phase space, 6 are
the spatial metric components  $g_{ij}$, 6 are their canonical momenta. Upon imposing four
constraints and three gauge symmetries, the phase space is described by five functions, and there is no symplectic
structure on this phase space.

Thus we need to impose new constraints derived from the Poisson brackets, but we
shall see that there are too many new constraints, thus it appears that no degree of freedom
is left, or the phase space is reduced to a smaller one still described by an odd number of fields,
this seems to us a fatal problem in Ho\v{r}ava's theory.

\section{Non-closure of constraints algebra and New constraints}

We start with the relativistic metric $g_{\mu\nu}$ in the usual ADM
decomposition,
\begin{equation}\label{}
    \pmatrix{ -N^2+N_iN^i & N_i  \cr N_i & g_{ij}}.
\end{equation}
for any theory of metric to be a theory of gravity, the lapse
function $N$ must be a function of both space and time, since
Newtonian potential is embedded in it. Even though Newtonian
potential can be included in the shift functions by choosing certain
special gauge, to obtain Newtonian equation which determines
Newtonian potential, the lapse function $N$ must be a function of
both space and time. Without Newtonian equation as a local
constraint, many unphysical solutions will emerge. In UV region the
 action in Ho\v{r}ava-Lifshitz theory takes the following form
\begin{equation}\label{LD}
    S=\frac{1}{16\pi G}\int dtd^3{\rm
    x}\sqrt{g}N\{(K_{ij}K^{ij}-\lambda
    K^2)-\frac{1}{k^4_W}C_{ij}C^{ij}\},
\end{equation}
where for simplicity we consider the first action proposed in
\cite{Horava2} in this section, we postpone a discussion of the
second action (which may be generated by flowing the above action
and flow to Einstein action in the IR region) in \cite{Horava2} to
sect.3, where similar calculations are carried out. The first two
terms in the above action comprise the kinetic term and the last
term is the potential term. This action satisfies the detailed
balance principle, since
$C^{ij}\equiv\epsilon^{ikl}\nabla_k(R^j_l-\frac{1}{4}R\delta^j_l)$
is the Cotton tensor and can be obtained from the variation of a
3-dimensional action $W[g_{ij}]$ \cite{Horava2}
\begin{equation}\label{}
    \sqrt{g}C^{ij}=\frac{\delta W}{\delta g_{ij}},
\end{equation}
where $W[g_{ij}]$ is the 3-dimensional Chern-Simons action.
$K_{ij}\equiv\frac{1}{2N}(\dot{g}_{ij}-\nabla_iN_j-\nabla_jN_i)$ are components of
the extrinsic curvature of the 3-dimensional hyper-surface with
constant $t$ and $K\equiv g^{ij}K_{ij}$ is its trace. $k_W$ is a
constant in this theory with mass dimension 1.

In the following, we focus on the case of $\lambda=1$ so
that the kinetic term in the action looks like that appearing in
Einstein-Hilbert action. Our calculation demonstrates that different
value of $\lambda$ will not change the final results
qualitatively, so we can make this choice.
Hereafter, we will
adopt the units in which $16\pi G=1$, for convenience. In the Hamiltonian formalism, we first compute the conjugate
momenta of $N, N_i$ and $g_{ij}$ denoted by $\pi, \pi^i$ and
$\pi^{ij}$, respectively. They have the explicit form
\begin{eqnarray}
 \label{cm1} \pi &=& \frac{\delta S}{\delta \dot{N}}=0 ,\\
 \label{cm2} \pi^i &=& \frac{\delta S}{\delta \dot{N}^i}=0, \\
  \label{cm3}\pi^{ij} &=&\frac{\delta S}{\delta \dot{g}_{ij}}=\sqrt{g}(K^{ij}-g^{ij}K).
\end{eqnarray}
The basic Poisson brackets of the canonical variables are
\begin{eqnarray}\label{}
   \{N(x),~\pi(y)\}_{\rm Pb}&=&\delta^3(\rm{x}-\rm{y}),\\
   \{N_i(x),~\pi^j(y)\}_{\rm Pb}&=&\delta^j_i\delta^3(\rm{x}-\rm{y}),\\
    \{g_{ij}(x),\pi^{kl}(y)\}_{\rm
    Pb}&=&\frac{1}{2}(\delta^k_i\delta^l_j+\delta^l_i\delta^k_j)\delta^3(\rm{x}-\rm{y}),
\end{eqnarray}
where we have adopted the convention of Poisson brackets between
canonical variables used by \cite{dewitt}. One should be aware of
the fact that $\pi_{ij}$ is not a tensor under coordinate
transformation, but behaves as a tensor density. In other words,
$\pi_{ij}/\sqrt{g}$ is a tensor. Then after performing a
Legendr\'{e} transformation, the Hamiltonian can be derived as
\begin{equation}\label{}
    H=\int\pi^{ij}\dot{g}_{ij}d^3{\rm x}-L=\int d^3{\rm x}(N\mathcal {H}+N^i\mathcal
    {H}_i),
\end{equation}
with $\mathcal {H}$ and $\mathcal{H}_i$ given by
\begin{eqnarray}
  \label{HC}\mathcal{H} &=& \frac{1}{\sqrt{g}}\pi^{ij}\mathcal {G}_{ijkl}\pi^{kl}+\frac{1}{k_W^4}\sqrt{g}C_{ij}C^{ij}, \\
 \label{MC} \mathcal{H}_i &=& -2g_{il}\partial_j\pi^{lj}-(2\partial_kg_{ij}-\partial_ig_{jk})\pi^{jk}.
\end{eqnarray}
where $\mathcal {G}_{ijkl}\equiv
\frac{1}{2}(g_{ik}g_{jl}+g_{il}g_{jk}-g_{ij}g_{kl})$ is the inverse
of de Witt metric. Since Eqs.(\ref{cm1}) and (\ref{cm2}) show that
the canonical momenta of $N$ and $N^i$ always vanish, the two
equations below are always satisfied
\begin{eqnarray}
  \dot{\pi} &=& \{\pi,~H\}_{\rm Pb}=\mathcal {H}, \\
   \dot{\pi}^i&=&\{\pi^i,~H\}_{\rm Pb}=\mathcal {H}_i,
\end{eqnarray}
provided the following constraints
\begin{eqnarray}
  \mathcal {H} &=& 0 ,\\
  \mathcal {H}_i&=& 0.
\end{eqnarray}
Above equations are usually called as the Hamiltonian constraint
and momentum constraint, respectively. Once the Poisson bracket is
added into the system as a part of structure, consistency requires
that the Poisson brackets among the constraints only generate
constraints. As discussed in the introduction, we anticipate that new constraints are generated
by the Poisson brackets among Hamiltonian constraints. To compute
the Poisson brackets of each pair of constraints, it is helpful to first
compute
\begin{eqnarray}
  \{\int d^3{\rm{x}^{\prime}}\zeta^k\mathcal {H}_k,g_{ij}\}_{\rm{Pb}} &=& -\zeta^k\partial_kg_{ij}-g_{jk}\partial_i\zeta^k-g_{ik}\partial_j\zeta^k, \\
   \{\int d^3{\rm{x}^{\prime}}\zeta^k\mathcal {H}_k,\pi^{ij}\}_{\rm{Pb}}&=&
   -\partial_k(\pi^{ij}\zeta^k)+\pi^{jk}\partial_k\zeta^i+\pi^{ik}\partial_k\zeta^j,
\end{eqnarray}
which reveal the role of $\mathcal {H}_i$'s as generators of
3-dimensional coordinate transformation, and
\begin{equation}\label{}
    \{\int d^3{\rm{x^{\prime}}}\eta\mathcal{H},~g_{ij}\}_{\rm{Pb}}=-\frac{\eta}{\sqrt{g}}(2g_{il}g_{jk}-g_{ij}g_{kl})\pi^{kl}=-2\eta
K_{ij}.
\end{equation}
Then one obtains the following two Poisson brackets
straightforwardly,
\begin{eqnarray}
    \label{hp1}\{\int d^3{\rm{x}}\zeta_1^i\mathcal{H}_i,~\int
    d^3{\rm{y}}\zeta_2^j\mathcal{H}_j\}_{\rm{Pb}}&=&\int
    d^3{\rm{x}}(\zeta_1^i\partial_i\zeta_2^k-\zeta_2^i\partial_i\zeta_1^k)\mathcal{H}_k,\\
   \label{hp2} \{\int d^3{\rm{x}}\zeta^i\mathcal{H}_i,~\int
    d^3{\rm{y}}\eta\mathcal{H}\}_{\rm{Pb}}&=&\int
    d^3{\rm{x}}\zeta^i\partial_i\eta\mathcal{H},
\end{eqnarray}
Eq.(\ref{hp1}) tells us that $\mathcal{H}_i$'s form a Lie algebra
corresponding to 3-dimensional diffeomorphism group. Eq.(\ref{hp2})
is a reflection of the fact that $\mathcal{H}$ behaves as a scalar
density under this transformation. Actually, one can expect these
results, since $\mathcal {H}_i$'s take the same form as in general
relativity. At this moment, no new constraints are generated, and we
have to compute the last Poisson bracket $\{\int
d^3{\rm{x}}\xi\mathcal{H},~\int
    d^3{\rm{y}}\eta\mathcal{H}\}_{\rm{Pb}}$.
This Poisson bracket
    is too complicated to analyze, instead, we consider a simple
    case that is $\sqrt{g}\xi=\delta^3({\rm x-y})$ with $\eta$ an
    arbitrary scalar function. In the process of computation, the following two formula are
    used,
    \begin{eqnarray}
     \delta R_{ij}&=&\frac{1}{2}g^{kl}(\nabla_l\nabla_j\delta g_{ki}+\nabla_l\nabla_i\delta g_{kj}-\nabla_l\nabla_k\delta g_{ij}-\nabla_i\nabla_j\delta g_{kl}),  \\
     \delta \Gamma^i_{jk}&=&\frac{1}{2}g^{il}(\nabla_j\delta g_{lk}+\nabla_k\delta g_{jl}-\nabla_l\delta
     g_{jk}).
    \end{eqnarray}
     After a tedious calculation, we obtain the following result
    expanded with respect to the covariant derivatives of $\eta$ in
    different orders.
\begin{equation}\label{PBHC}
\{\mathcal{H}({\rm x}),~\int
    d^3{\rm{y}}\eta\mathcal{H}\}_{\rm{Pb}}=-2\sqrt{g}\frac{1}{k^4_W}(\alpha^{ijk}\nabla_k\nabla_j\nabla_i\eta+\beta^{ij}\nabla_j\nabla_i\eta+\gamma^i\nabla_i\eta+\omega\eta),
\end{equation}
with $\alpha^{ijk}$ given by
\begin{equation}\label{}
    \alpha^{ijk}
    =(\widetilde{C}^{ilm}g^{jk}+\widetilde{C}^{klm}g^{ij}-\widetilde{C}^{ilk}g^{jm}-\widetilde{C}^{kli}g^{jm})K_{lm},
\end{equation}
where $\widetilde{C}^{ijk}$ is defined  as $\epsilon^{ijl}C_l^{~k}$,
in which $C_l^{~k}=g_{lm}C^{mk}$ with $C^{mk}$ the Cotton tensor
defined before. The term including the second order covariant
derivative of $\eta$ takes the form
\begin{equation}\label{}
    \beta^{ij}\nabla_j\nabla_i\eta=\nabla_{(j}\nabla_i\eta\nabla_{k)_{\rm
    c}}(K_{lm}\widetilde{C}^{ilm}g^{jk}-K_{lm}\widetilde{C}^{ilk}g^{jm}),
\end{equation}
where the notation $(kji)_c\equiv(ijk+jki+kij)$ represents the
cyclic permutation among indices $i,j,k$. The term proportional to
the first order derivative of $\eta$ is a little lengthy, and we have
to introduce some abbreviations
\begin{eqnarray}
   t^{ijklm}&=&\widetilde{C}^{ijm}g^{kl}+\widetilde{C}^{imj}g^{kl}+\widetilde{C}^{jml}g^{ik},  \\
   s^{ijklm}&=&\widetilde{C}^{ijm}g^{kl}+\widetilde{C}^{imj}g^{kl}-\widetilde{C}^{jml}g^{ik}.
\end{eqnarray}
Then
\begin{eqnarray}
\nonumber  \gamma^i\nabla_i\eta &=& t^{mlkji}\nabla_{(i}\eta\nabla_m\nabla_{k)_{\rm c}}K_{jl}+K_{jl}\nabla_{(i}\eta \nabla_k\nabla_{m)_{\rm c}}t^{mlkji} \\
\nonumber   &&+2s^{lmijk}\nabla_i\eta\nabla_{[l}\nabla_{k]}K_{jm}+2K_{jm}\nabla_i\eta\nabla_{[k}\nabla_{l]}s^{lmijk}\\
\nonumber &&+2(\widetilde{C}^{klj}R^i_l+\widetilde{C}^{lki}R^j_l+\widetilde{C}^{lij}R^k_l)K_{jk}\nabla_i\eta\\
   &&+(\frac{1}{2}R^i_{jkl}\widetilde{C}^{klj}-R^i_{jkl}\widetilde{C}^{jkl})K\nabla_i\eta
\end{eqnarray}
The last one $\omega$ is
\begin{eqnarray}
\nonumber  \omega &=& \nabla_i(\widetilde{C}^{jkl}R^i_kK_{jl}+\widetilde{C}^{jik}R^l_jK_{kl}+\widetilde{C}^{kji}R^l_kK_{jl}) \\
   \nonumber&&+\widetilde{C}^{ijk}(\nabla_i\nabla_l\nabla_kK^l_{~j}+\nabla_i\nabla_l\nabla_jK^l_{~k}-\nabla_i\nabla^l\nabla_lK_{jk}-\nabla_i\nabla_k\nabla_jK)\\
   &&+(K^l_{~j}\nabla_k\nabla_l\nabla_i+K^l_{~k}\nabla_j\nabla_l\nabla_i-K_{jk}\nabla^l\nabla_l\nabla_i-K\nabla_j\nabla_k\nabla_i)\widetilde{C}^{ijk}.
\end{eqnarray}

At first sight, one may think that constraints from the third covariant derivatives of $\eta$ are
 $\alpha^{ijk}$=0, it is not true.
The reason is the following. We first notice that
\begin{equation}\label{eta1}
     \nabla_k\nabla_j\nabla_i\eta=\nabla_k\nabla_i\nabla_j\eta.
\end{equation}
This originates from the formula of second order covariant
derivatives of a scalar function, which is
\begin{equation}\label{}
   \nabla_j\nabla_i\eta=\partial_j\partial_i\eta-\Gamma^k_{ij}\partial_k\eta,
\end{equation}
where $\Gamma^k_{ij}$ is the Christopher symbol. For a torsionless
space, $\Gamma^k_{ij}$ is symmetric in the two lower indices,
combining with commutativity of partial derivatives, we have Eq.(\ref{eta1}). Furthermore, we have the following
identities,
\begin{equation}\label{eta2}
    \nabla_k\nabla_j\nabla_i\eta=\frac{1}{3}\nabla_{(k}\nabla_j\nabla_{i)_c}\eta+\frac{1}{3}(\nabla_{k}\nabla_i\nabla_{j}-\nabla_{i}\nabla_k\nabla_{j})\eta+\frac{1}{3}(\nabla_{k}\nabla_j\nabla_{i}-\nabla_{j}\nabla_k\nabla_{i})\eta,
\end{equation}
where we used Eq.(\ref{eta1}). With the help of Eq.(\ref{eta1}), we find that the first term in above equation is
completely symmetric in three indices. If we denote the
symmetrization among three indices by
$(ijk)\equiv\frac{1}{6}(ijk+jki+kij+jik+kji+ikj)$, then
$\frac{1}{3}\nabla_{(k}\nabla_j\nabla_{i)_c}\eta=\nabla_{(k}\nabla_j\nabla_{i)}\eta$.
The remaining two terms in Eq.(\ref{eta2}) can be reduced to the
first order derivative of $\eta$ by utilizing the following formula,
\begin{equation}\label{}
(\nabla_{k}\nabla_i\nabla_{j}-\nabla_{i}\nabla_k\nabla_{j})\eta=R^l_{jik}\nabla_l\eta.
\end{equation}
After this process, it is clear that the
third order covariant derivative term in Eq.(\ref{PBHC}) can be written as
\begin{equation}\label{}
    \alpha^{ijk}\nabla_{(k}\nabla_j\nabla_{i)}\eta.
\end{equation}
Then inheriting the symmetry of
$\nabla_{(k}\nabla_j\nabla_{i)}\eta$, the effective components of
$\alpha^{ijk}$ are $\alpha^{(ijk)}$. Explicitly, they take the
following form
\begin{eqnarray}\label{}
\nonumber \alpha^{(ijk)}=\frac{2}{3}(\widetilde{C}^{klm}g^{ij}&+&\widetilde{C}^{jlm}g^{ik}+\widetilde{C}^{ilm}g^{jk})K_{ml}-\frac{1}{3}(\widetilde{C}^{ilk}g^{jm}+\widetilde{C}^{kli}g^{jm}+\widetilde{C}^{jli}g^{km})K_{ml}\\
                        &&-\frac{1}{3}(\widetilde{C}^{ilj}g^{km}+\widetilde{C}^{klj}g^{im}+\widetilde{C}^{jlk}g^{im})K_{ml}.
\end{eqnarray}
Because  covariant derivatives of $\eta$ of different order are
independent, consistency requires that the coefficients in front of
covariant derivatives of $\eta$ should vanish (Because $\mathcal
{H}$ is a first class constraint and the reason will be given later). We deduce from the term
$\alpha^{ijk}\nabla_{(k}\nabla_j\nabla_{i)}\eta$ that
\begin{equation}\label{NC}
\alpha^{(ijk)}=0.
\end{equation}
To see whether this condition gives rise to new constraints, we will
work in a special frame where Eq.(\ref{NC}) becomes simple. First
we write the 3-dimensional space metric tensor $g_{ij}dx^idx^j$ in
terms of vielbein, namely
\begin{equation}\label{}
    g_{ij}dx^idx^j=\delta_{ab}\theta^a\theta^b,
\end{equation}
where $\delta_{ab}$ is the Kronecker delta function, new basis
$\theta^a$ is related to $dx^i$ by $\theta^a=\theta^a_idx^i$. Since
Cotton tensor is a symmetric tensor, we can diagonalize it through
an orthogonal transformation $\mathcal {O}(\mbox{x})$ at each given
point. So in terms of new basis defined by
$\tilde{\theta}^\alpha=\mathcal {O}^{\alpha}_{~b}\theta^b$, the
component of Cotton tensor becomes
\begin{equation}\label{}
C^{\alpha\beta}=(\mathcal {O}C\mathcal
{O}^T)^{\alpha\beta}=\mbox{diag}\{C_1,~C_2,~-C_1-C_2\},
\end{equation}
where $C_1, C_2, \mbox{and} -C_1-C_2$ are three eigenvalues of
Cotton tensor, where we used the property that Cotton tensor is
traceless. Contracting two sides of Eq.(\ref{cm3}), we obtain
\begin{equation}\label{}
    \sqrt{g}K=-\frac{\pi}{2}~~~~~(\pi=g^{ij}\pi_{ij}),
\end{equation}
this helps us to reexpress $K_{ij}$ by $\pi^{ij}$
\begin{equation}\label{}
    \sqrt{g}K^{ij}=\pi^{ij}-\frac{1}{2}g^{ij}\pi.
\end{equation}

Now, we find that Eq.(\ref{NC}) gives the following seven independent equations
\begin{equation}\label{}
C_1\pi^{12}= 0\, ,\\~~~~~~~~C_2\pi^{12}=0,
\end{equation}
\begin{equation}\label{}
    C_1\pi^{13}=0\, ,\\~~~~~~~~C_2\pi^{13}=0,
\end{equation}
\begin{equation}\label{}
    C_1\pi^{23} =0\, ,\\~~~~~~~~C_2\pi^{23}=0,
\end{equation}
\begin{equation}\label{}
    (C_1+2C_2)\pi^{11}-(2C_1+C_2)\pi^{22}+(C_1-C_2)\pi^{33}=0.
\end{equation}
The solutions of these constraints can be separated into two classes
\begin{itemize}
 \item 1) $C_1=C_2=0$ or in other words Cotton tensor should
           vanish. We notice that the other coefficient tensors also vanish because they all consist of Cotton tensor and its covariant derivatives.
           Since the fact that a tensor vanishes is a coordinate independent statement, these constraints are
           still required to be satisfied, even after one uses the three coordinate
           transformations to fix three components of
           $g_{ij}$. In a special frame, we see that the vanishing of Cotton tensor gives two constraint equations,
           actually, in a general coordinate frame, the vanishing of Cotton tensor indeed provides only two constraints. The reason
           is that the following   properties of Cotton tensor
           \begin{equation}\label{}
            C_{ij}=C_{ji},~~~~g^{ij}C_{ij}=0,~~~~\nabla_jC^{ij}=0,
           \end{equation}
           make Cotton tensor itself have only two independent components.
           Gauge-fixing  and vanishing of
           Cotton tensor altogether make only one degree of freedom in $g_{ij}$
           be physical. Now,
           the Hamiltonian constraint and three momentum constraints eliminate four conjugate momenta, leaving two components of $\pi^{ij}$
           be physical. Altogether, the phase space is described by three unpaired fields.
 \item 2)
 \begin{equation}\label{}
    C_1\neq0~~
    \mbox{or}~~C_2\neq0,~~\pi^{12}=0,~~\pi^{13}=0,~~\pi^{23}=0.
 \end{equation}
  This case is rather bad, it indicates that all the
conjugate momenta of $g_{ij}$ are unphysical, since three momentum
constraints already elminates three of the six conjugate momenta.
\end{itemize}
To interpret above results with the approach given by Dirac , we
consider the time derivative of the constraint. Utilizing
Eqs.(\ref{hp1}), (\ref{hp2}) and (\ref{PBHC}), we obtain
\begin{eqnarray}
  \frac{d\mathcal {H}_i}{dt} &=& \partial_i\mathcal {N}\mathcal {H}+(\partial_i\mathcal {N}^j)\mathcal {H}_j+\partial_j(\mathcal {N}^j\mathcal {H}_i), \\
  \frac{d\mathcal {H}}{dt} &=&  (\partial_i\mathcal {N}^i)\mathcal
  {H}+\triangle\mathcal {N},
\end{eqnarray}
where the operator $\triangle$ is defined as
\begin{equation}\label{}
    \triangle=
    -2\sqrt{g}\frac{1}{k^4_W}(\alpha^{ijk}\nabla_k\nabla_j\nabla_i+\beta^{ij}\nabla_j\nabla_i+\gamma^i\nabla_i+\omega),
\end{equation}
where the coefficients $\alpha^{ijk}$, $\beta^{ij}$, $\gamma^i$ and
$\omega$ take the same expression as in Eq.(\ref{PBHC}). The
preservation of constraint in time require $\frac{d\mathcal
{H}_i}{dt}\simeq0,~\frac{d\mathcal {H}}{dt}\simeq0$ on the
constrained phase space. The first one is satisfied because
$\mathcal {H}\simeq0$ and $\mathcal {H}_i\simeq0$. While to satisfy
the second one gives a differential equation $\triangle\mathcal
{N}\simeq0$. It is remarkable that $\triangle$ has no inverse on the
whole constrained phase space, due to the existence of configuration
$g_{ij}$, $\pi^{ij}$ making the coefficients in front of covariant
derivative vanish. As we have analyzed before, the configuration
$g_{ij}$ with $C_{ij}=0$ can achieve this. Therefore, $\mathcal {H}$
cannot be perceived as a second class constraint, since to define
the Dirac bracket associated with the second class constraint, the
inverse of $\triangle$ is indispensable. So $\mathcal {H}$ can only
be a first class constraint, and its Poisson bracket generates
constraints according to the property of first class constraint.
This justifies our previous treatment of the terms yielded by
$\{\mathcal{H}({\rm x}),~\int
    d^3{\rm{y}}\eta\mathcal{H}\}_{\rm{Pb}}$.

To summarize, we have found new
constraints generated from the Poisson brackets of Hamiltonian
constraint. These new constraints reduce further the phase space in
a way that it appears that no symplectic structure exists, or
eliminate all the degrees of freedom. We expect that the Poisson
brackets among the new constraints and $\mathcal {H}$, $\mathcal
{H}_i$ yield more constraints, until the constraints form a closed
algebra. When this is done, all the constraints are called the first
class. Note that the Hamiltonian density is not a second class
constraint, since the Poisson bracket obviously does not have an
inverse.

Put together, all new constraints either eliminate all degrees of freedom, or make the
reduced phase space unphysical.

A complete discussion deserves to be carried out in another work.

\section{Discussion and conclusion}
In previous section, our calculation shows explicitly that $\mathcal
{H}$ and $\mathcal {H}_i$ do not form a closed algebra as what
happens in general relativity. Intuitively, we feel that
non-closure of the Poisson brackets among $\mathcal {H}$ and $\mathcal
{H}_i$ can be interpreted by considering the relationship between
constraints and gauged symmetry . The details are what follows. One
can calculate the Poisson bracket between the combined constraint
$N\mathcal{H}+N^i\mathcal{H}_i$ and $g_{ij}$ then obtains
\begin{equation}\label{hp3}
\{\int
d^3{\rm{x}}\eta(N\mathcal{H}+\mathcal{N}^i\mathcal{H}_i),~g_{ij}\}_{\rm{Pb}}=-\partial_i\eta
N_j-\partial_j\eta N_i-\eta \dot{g}_{ij}.
\end{equation}
It is nothing but the variation of $g_{ij}$ under (3+1)-dimensional
coordinate transformation in time direction. So the four generators
of (3+1)-dimensional coordinate transformation are $(N\mathcal{H}+N^i\mathcal{H}_i)$ and $\mathcal{H}_i$. This
seems to suggest that the constraints $\mathcal{H}$ and
$\mathcal{H}_i$ require the theory to have full
(3+1)-dimensional covariance to respect them, or more
constraints should be added. The reason is that in a field theory,
a local constraint is always accompanied by a gauge symmetry. if we
denote the constraint by $\mathcal {C}$, the meaning of constraint
can be expressed by
\begin{equation}\label{c2}
    \mathcal {C}|{\rm phys}\rangle\approx0.
\end{equation}
This equation implies that the physical state is gauge invariant if
there is a corresponding symmetry. In H-L theory, although there is
no diffeomorphism invariance in the time direction, Eq.(\ref{hp3})
still looks like such a transformation and imposing this constraint
is somewhat in conflicts of the starting point\footnote{Imposing
local Hamiltonian constraint, but lacking the diffeomorphism
invariance in the time direction may cause strong coupling problem
\cite{Niz} in the IR region of H-L Theory. Before the revised
version of our paper appears, paper \cite{Mokoh} also found this
independently.}.

What we have discussed previously is based on the UV action of H-L
theory. To complete our discussion, the full action
containing the description of H-L theory both in UV and IR region
should be taken into account. In the IR region, some operators
with lower mass dimension will become relevant. The detailed balance
principle forces the action to take the following form \cite{Horava2}
\begin{eqnarray}
\nonumber  S &=& \int dtd^3\mbox{x}\sqrt{g}N\{(K_{ij}K^{ij}-\lambda K^2)-\frac{1}{k^4_W}C_{ij}C^{ij} \\
\nonumber    && +\frac{\mu}{k^2_W}\epsilon^{ijk}R_{il}\nabla_jR^l_k-\frac{\mu^2}{4}R_{ij}R^{ij} \\
          &&
          +\frac{\mu^2}{4(1-3\lambda)}(\frac{1-4\lambda}{4}R^2+\Lambda_WR-3\Lambda^2_W)\},
\end{eqnarray}
For this action to be a deformation of Einstein-Hilbert action in
the IR region, it is natural to express this action in relativistic
coordinates by rescaling $t$,
\begin{equation}\label{}
    x_0=ct,
\end{equation}
with the emergent speed of light and effective cosmological constant
given by
\begin{equation}\label{}
    c=\frac{\mu}{2}\sqrt{\frac{\Lambda_W}{1-3\lambda}},~~~~~\Lambda=\frac{3}{2}\Lambda_W.
\end{equation}
At the same time, parameter $\lambda$  should be equal to 1 for
the kinetic term taking the same form as its counterpart in general
relativity. Thus to have a real speed of light, $\Lambda_W$ should
be negative. Setting
$\lambda=1$, the action becomes
\begin{eqnarray}
\nonumber S &=&  \int dtd^3\mbox{x}\sqrt{g}N\{(K_{ij}K^{ij}- K^2)-\frac{1}{k^4_W}C_{ij}C^{ij} \\
\nonumber    && +\frac{\mu}{k^2_W}\epsilon^{ijk}R_{il}\nabla_jR^l_k-\frac{\mu^2}{4}R_{ij}R^{ij} \\
          &&
          +\frac{\mu^2}{8}(\frac{3}{4}R^2-\Lambda_WR+3\Lambda^2_W)\},
\end{eqnarray}

The  momentum constraints corresponding to this action remain the
same form as in Eq.(\ref{MC}), the Hamiltonian density contains more
terms than in Eq.(\ref{HC}). We  denote the new Hamiltonian
constraint by $\widetilde{\mathcal {H}}$ to distinguish it from the
 previous Hamiltonian constraint
\begin{eqnarray}\label{}
\nonumber    \widetilde{\mathcal {H}}/\sqrt{g}=&&
\frac{1}{g}\pi^{ij}\mathcal
    {G}_{ijkl}\pi^{kl}-R+2\Lambda+\frac{1}{k_W^4}C_{ij}C^{ij}-\frac{\mu}{k^2_W}\epsilon^{ijk}R_{il}\nabla_jR^l_k+\frac{\mu^2}{4}R_{ij}R^{ij}\\
          &&~~~~~~-\frac{3\mu^2}{32}R^2,
\end{eqnarray}
where the speed of light has been set to 1 and then the relation
$\Lambda_W=-{8}/{\mu^2}$ has been used. It is noticed that the first
three terms make up of Hamiltonian constraint in general relativity.
Calculation of Poisson bracket of $\widetilde{\mathcal {H}}$ is
similar to the previous one, and we obtain the following result,
\begin{eqnarray}\label{}
\nonumber\{\widetilde{\mathcal{H}}({\rm x}),~\int
    d^3{\rm{y}}\eta\widetilde{\mathcal{H}}\}_{\rm{Pb}}&=&-2\sqrt{g}\frac{1}{k^4_W}(\widetilde{\alpha}^{ijk}\nabla_k\nabla_j\nabla_i\eta+\widetilde{\beta}^{ij}\nabla_j\nabla_i\eta+\widetilde{\gamma}^i\nabla_i\eta+\widetilde{\omega}\eta),\\
    &&+\partial_i(\eta g^{ij}\mathcal {H}_i)-g^{ij}\mathcal
    {H}_i\partial_j\eta.
\end{eqnarray}
The last two terms come from the well known result of Poisson
bracket between the Hamiltonian constraint of general relativity. In the
previous section, our discussion mainly concentrates on coefficient
tensor of the third order covariant derivatives of $\eta$, we want
to see whether this new coefficient tensor will change our results.
Then we compute $\widetilde{\alpha}^{ijk}$ and find that it amounts
to replacing the Cotton tensor $C^{ij}$ in $\alpha^{ijk}$ by
$C^{ij}-\frac{{\mu}{k^2_W}}{2}R^{ij}$. The new contribution
$-\frac{{\mu}{k^2_W}}{2}R^{ij}$ comes from term
$-\frac{\mu}{k^2_W}\epsilon^{ijk}R_{il}\nabla_jR^l_k $, since it
also contains the third order derivative of metric $g_{ij}$. If we
denote the tensor $C^{ij}-\frac{{\mu}{k^2_W}}{2}R^{ij}$ by
$\Sigma^{ij}$, then the vanishing of $\widetilde{\alpha}^{(ijk)}$
becomes the following seven independent constraints,
\begin{equation}\label{}
(\Sigma_1-\Sigma_2)\pi^{12}= 0\,
,\\~~~~~~~~(\Sigma_2-\Sigma_3)\pi^{12}=0,
\end{equation}
\begin{equation}\label{}
    (\Sigma_1-\Sigma_2)\pi^{13}=0\, ,\\~~~~~~~~(\Sigma_2-\Sigma_3)\pi^{13}=0,
\end{equation}
\begin{equation}\label{}
    (\Sigma_1-\Sigma_2)\pi^{23} =0\, ,\\~~~~~~~~(\Sigma_2-\Sigma_3)\pi^{23}=0,
\end{equation}
\begin{equation}\label{}
    (\Sigma_2-\Sigma_3)\pi^{11}+(\Sigma_3-\Sigma_1)\pi^{22}+(\Sigma_1-\Sigma_2)\pi^{33}=0.
\end{equation} To obtain above equations, we also work in basis
where $\Sigma_{ij}$ is diagonalized by
$\Sigma_{ij}=\mbox{diag}\{\Sigma_1,~\Sigma_2,~\Sigma_3\}$.
the three eigenvalues are independent
because usually $\Sigma^{ij}$ does not satisfy the traceless
condition. Despite this difference, our analysis can still be
applied to this case and our conclusion is unchanged, the same difficulty remains.

One may choose not to impose the Hamiltonian constraint, this contradicts the requirement
that the lapse function is a full-fledged function. Moreover, if one does not impose this constraint
at the beginning, how
can one obtain the usual Hamiltonian constraint in Einstein theory in the infrared regime?
(This constraint is crucial in going back to the Newtonian limit)
Finally, we want to remark that even if one can come up with some cure of this
problem, it will be very difficult to come up with modified theory containing a spin
2 graviton.

\emph{Note added}: When this paper was reviewed, we were informed
that another paper \cite{Blas} holds a different opinion on the
constraint structure in Ho\v{r}ava-Lifshitz theory. We disagree with
the result in \cite{Blas} for the following reason. Most of
discussion in that work is based on a perturbative method, while
this method is not suitable for discussing the fundamental degrees
of freedom of a theory. For instance, the new constraints obtained
in our paper are at least of the second order perturbations around
Minkowski background (adopted in \cite{Blas}), so at the linear
level they do not show up. However, these constraints already
determine how many degrees of freedom a theory can have before one
carries out a perturbative calculation. In summary, our point of
view is that this perturbative method may be useful in solving
equations of motion but is invalid for counting the number of
degrees of freedom of a theory.

\section*{Acknowledgments} We thank Wei Song and Yong-Shi Wu  for useful discussions. This work was supported by
a NSFC Grant No. 10535060/A050207, a NSFC group Grant No. 10821504,
and a 973 project Grant No. 2007CB815401.

\end{document}